\catcode`\@=11					



\font\fiverm=cmr5				
\font\fivemi=cmmi5				
\font\fivesy=cmsy5				
\font\fivebf=cmbx5				

\skewchar\fivemi='177
\skewchar\fivesy='60


\font\sixrm=cmr6				
\font\sixi=cmmi6				
\font\sixsy=cmsy6				
\font\sixbf=cmbx6				

\skewchar\sixi='177
\skewchar\sixsy='60


\font\sevenrm=cmr7				
\font\seveni=cmmi7				
\font\sevensy=cmsy7				
\font\sevenit=cmti7				
\font\sevenbf=cmbx7				

\skewchar\seveni='177
\skewchar\sevensy='60


\font\eightrm=cmr8				
\font\eighti=cmmi8				
\font\eightsy=cmsy8				
\font\eightit=cmti8				
\font\eightbf=cmbx8				

\skewchar\eighti='177
\skewchar\eightsy='60


\font\ninei=cmmi9
\font\ninesy=cmsy9

\skewchar\ninei='177
\skewchar\ninesy='60


\font\tenrm=cmr10				
\font\teni=cmmi10				
\font\tensy=cmsy10				
\font\tenex=cmex10				
\font\tenit=cmti10				
\font\tensl=cmsl10				
\font\tenbf=cmbx10				
\font\tentt=cmtt10				
\font\tenss=cmss10				
\font\tensc=cmcsc10				
\font\tenbi=cmmib10				

\skewchar\teni='177
\skewchar\tenbi='177
\skewchar\tensy='60

\def\tenpoint{\ifmmode\err@badsizechange\else
	\textfont0=\tenrm \scriptfont0=\sevenrm \scriptscriptfont0=\fiverm
	\textfont1=\teni  \scriptfont1=\seveni  \scriptscriptfont1=\fivemi
	\textfont2=\tensy \scriptfont2=\sevensy \scriptscriptfont2=\fivesy
	\textfont3=\tenex \scriptfont3=\tenex   \scriptscriptfont3=\tenex
	\textfont4=\tenit \scriptfont4=\sevenit \scriptscriptfont4=\sevenit
	\textfont5=\tensl
	\textfont6=\tenbf \scriptfont6=\sevenbf \scriptscriptfont6=\fivebf
	\textfont7=\tentt
	\textfont8=\tenbi \scriptfont8=\seveni  \scriptscriptfont8=\fivemi
	\def\rm{\tenrm\fam=0 }%
	\def\it{\tenit\fam=4 }%
	\def\sl{\tensl\fam=5 }%
	\def\bf{\tenbf\fam=6 }%
	\def\tt{\tentt\fam=7 }%
	\def\ss{\tenss}%
	\def\sc{\tensc}%
	\def\bmit{\fam=8 }%
	\rm\setparameters\setbaselines\fi}


\font\twelverm=cmr12				
\font\twelvei=cmmi12				
\font\twelvesy=cmsy10	scaled\magstep1		
\font\twelveex=cmex10	scaled\magstep1		
\font\twelveit=cmti12				
\font\twelvesl=cmsl12				
\font\twelvebf=cmbx12				
\font\twelvett=cmtt12				
\font\twelvess=cmss12				
\font\twelvesc=cmcsc10	scaled\magstep1		
\font\twelvebi=cmmib10	scaled\magstep1		

\skewchar\twelvei='177
\skewchar\twelvebi='177
\skewchar\twelvesy='60

\def\twelvepoint{\ifmmode\err@badsizechange\else
	\textfont0=\twelverm \scriptfont0=\eightrm \scriptscriptfont0=\sixrm
	\textfont1=\twelvei  \scriptfont1=\eighti  \scriptscriptfont1=\sixi
	\textfont2=\twelvesy \scriptfont2=\eightsy \scriptscriptfont2=\sixsy
	\textfont3=\twelveex \scriptfont3=\tenex   \scriptscriptfont3=\tenex
	\textfont4=\twelveit \scriptfont4=\eightit \scriptscriptfont4=\sevenit
	\textfont5=\twelvesl
	\textfont6=\twelvebf \scriptfont6=\eightbf \scriptscriptfont6=\sixbf
	\textfont7=\twelvett
	\textfont8=\twelvebi \scriptfont8=\eighti  \scriptscriptfont8=\sixi
	\def\rm{\twelverm\fam=0 }%
	\def\it{\twelveit\fam=4 }%
	\def\sl{\twelvesl\fam=5 }%
	\def\bf{\twelvebf\fam=6 }%
	\def\tt{\twelvett\fam=7 }%
	\def\ss{\twelvess}%
	\def\sc{\twelvesc}%
	\def\bmit{\fam=8 }%
	\rm\setparameters\setbaselines\fi}


\font\fourteenrm=cmr12	scaled\magstep1		
\font\fourteeni=cmmi12	scaled\magstep1		
\font\fourteensy=cmsy10	scaled\magstep2		
\font\fourteenex=cmex10	scaled\magstep2		
\font\fourteenit=cmti12	scaled\magstep1		
\font\fourteensl=cmsl12	scaled\magstep1		
\font\fourteenbf=cmbx12	scaled\magstep1		
\font\fourteentt=cmtt12	scaled\magstep1		
\font\fourteenss=cmss12	scaled\magstep1		
\font\fourteensc=cmcsc10 scaled\magstep2	
\font\fourteenbi=cmmib10 scaled\magstep2	

\skewchar\fourteeni='177
\skewchar\fourteenbi='177
\skewchar\fourteensy='60

\def\fourteenpoint{\ifmmode\err@badsizechange\else
	\textfont0=\fourteenrm \scriptfont0=\tenrm \scriptscriptfont0=\sevenrm
	\textfont1=\fourteeni  \scriptfont1=\teni  \scriptscriptfont1=\seveni
	\textfont2=\fourteensy \scriptfont2=\tensy \scriptscriptfont2=\sevensy
	\textfont3=\fourteenex \scriptfont3=\tenex \scriptscriptfont3=\tenex
	\textfont4=\fourteenit \scriptfont4=\tenit \scriptscriptfont4=\sevenit
	\textfont5=\fourteensl
	\textfont6=\fourteenbf \scriptfont6=\tenbf \scriptscriptfont6=\sevenbf
	\textfont7=\fourteentt
	\textfont8=\fourteenbi \scriptfont8=\tenbi \scriptscriptfont8=\seveni
	\def\rm{\fourteenrm\fam=0 }%
	\def\it{\fourteenit\fam=4 }%
	\def\sl{\fourteensl\fam=5 }%
	\def\bf{\fourteenbf\fam=6 }%
	\def\tt{\fourteentt\fam=7}%
	\def\ss{\fourteenss}%
	\def\sc{\fourteensc}%
	\def\bmit{\fam=8 }%
	\rm\setparameters\setbaselines\fi}


\font\seventeenrm=cmr10 scaled\magstep3		


\newdimen\rp@
\newcount\@basestretchnum
\newskip\@baseskip
\newskip\headskip
\newskip\footskip


\def\setparameters{\rp@=.1em
	\headskip=24\rp@
	\footskip=\headskip
	\delimitershortfall=5\rp@
	\nulldelimiterspace=1.2\rp@
	\scriptspace=0.5\rp@
	\abovedisplayskip=10\rp@ plus3\rp@ minus5\rp@
	\belowdisplayskip=10\rp@ plus3\rp@ minus5\rp@
	\abovedisplayshortskip=5\rp@ plus2\rp@ minus4\rp@
	\belowdisplayshortskip=10\rp@ plus3\rp@ minus5\rp@
	\normallineskip=\rp@
	\lineskip=\normallineskip
	\normallineskiplimit=0pt
	\lineskiplimit=\normallineskiplimit
	\jot=3\rp@
	\setbox0=\hbox{\the\textfont3 B}\p@renwd=\wd0
	\skip\footins=12\rp@ plus3\rp@ minus3\rp@
	\skip\topins=0pt plus0pt minus0pt}


\def\setbaselines{\maxdepth=4\rp@\baselinestretch=\@basestretchnum}


\def\baselinestretch{\afterassignment\@basestretch\@basestretchnum}
\def\@basestretch{%
	\@baseskip=12\rp@ \divide\@baseskip by1000
	\normalbaselineskip=\@basestretchnum\@baseskip
	\baselineskip=\normalbaselineskip
	\bigskipamount=\the\baselineskip
		plus.25\baselineskip minus.25\baselineskip
	\medskipamount=.5\baselineskip
		plus.125\baselineskip minus.125\baselineskip
	\smallskipamount=.25\baselineskip
		plus.0625\baselineskip minus.0625\baselineskip
	\setbox\strutbox=\hbox{\vrule height.708\baselineskip
		depth.292\baselineskip width0pt }}



\def\makeheadline{\vbox to0pt{\baselinestretch=1000
	\vskip-\headskip \vskip1.5pt
	\line{\vbox to\ht\strutbox{}\the\headline}\vss}\nointerlineskip}

\def\makefootline{\baselineskip=\footskip\line{\the\footline}}

\def\big#1{{\hbox{$\left#1\vbox to8.5\rp@ {}\right.\n@space$}}}
\def\Big#1{{\hbox{$\left#1\vbox to11.5\rp@ {}\right.\n@space$}}}
\def\bigg#1{{\hbox{$\left#1\vbox to14.5\rp@ {}\right.\n@space$}}}
\def\Bigg#1{{\hbox{$\left#1\vbox to17.5\rp@ {}\right.\n@space$}}}


\mathchardef\alpha="710B
\mathchardef\beta="710C
\mathchardef\gamma="710D
\mathchardef\delta="710E
\mathchardef\epsilon="710F
\mathchardef\zeta="7110
\mathchardef\eta="7111
\mathchardef\theta="7112
\mathchardef\iota="7113
\mathchardef\kappa="7114
\mathchardef\lambda="7115
\mathchardef\mu="7116
\mathchardef\nu="7117
\mathchardef\xi="7118
\mathchardef\pi="7119
\mathchardef\rho="711A
\mathchardef\sigma="711B
\mathchardef\tau="711C
\mathchardef\upsilon="711D
\mathchardef\phi="711E
\mathchardef\chi="711F
\mathchardef\psi="7120
\mathchardef\omega="7121
\mathchardef\varepsilon="7122
\mathchardef\vartheta="7123
\mathchardef\varpi="7124
\mathchardef\varrho="7125
\mathchardef\varsigma="7126
\mathchardef\varphi="7127
\mathchardef\imath="717B
\mathchardef\jmath="717C
\mathchardef\ell="7160
\mathchardef\wp="717D
\mathchardef\partial="7140
\mathchardef\flat="715B
\mathchardef\natural="715C
\mathchardef\sharp="715D


\def\err@badsizechange{%
	\immediate\write16{--> Size change not allowed in math mode, ignored}}

\baselinestretch=1000
\tenpoint

\catcode`\@=12					
\catcode`\@=11
\expandafter\ifx\csname @iasmacros\endcsname\relax
	\global\let\@iasmacros=\par
\else	\immediate\write16{}
	\immediate\write16{Warning:}
	\immediate\write16{You have tried to input iasmacros more than once.}
	\immediate\write16{}
	\endinput
\fi
\catcode`\@=12


\def\rmb{\seventeenrm}

\def\singlespace{\baselineskip=\normalbaselineskip}
\def\halfspace{\baselineskip=1.5\normalbaselineskip}
\def\doublespace{\baselineskip=2\normalbaselineskip}


\def\AB{\bigskip\parindent=40pt
        \centerline{\bf ABSTRACT}\medskip\halfspace\narrower}
\def\AE{\bigskip\nonarrower\doublespace}
\def\nonarrower{\advance\leftskip by-\parindent
	\advance\rightskip by-\parindent}


\def\boxit#1{\vbox{\hrule\hbox{\vrule\kern3pt
	\vbox{\kern3pt#1\kern3pt}\kern3pt\vrule}\hrule}}

\def\hence{\leavevmode\hbox{\bf .\raise5.5pt\hbox{.}.} }

\def\square{\mathord{\dalemb{5.9}{6}\hbox{\hskip1pt}}}
\def\dalemb#1#2{{\vbox{\hrule height.#2pt
	\hbox{\vrule width.#2pt height#1pt \kern#1pt \vrule width.#2pt}
	\hrule height.#2pt}}}
\def\gtorder{\mathrel{\raise.3ex\hbox{$>$}\mkern-14mu
             \lower0.6ex\hbox{$\sim$}}}
\def\ltorder{\mathrel{\raise.3ex\hbox{$<$}\mkern-14mu
             \lower0.6ex\hbox{$\sim$}}}

\newdimen\fullhsize
\newbox\leftcolumn
\def\twoup{\hoffset=-.5in \voffset=-.25in
  \hsize=4.75in \fullhsize=10in \vsize=6.9in
  \def\fullline{\hbox to\fullhsize}
  \let\lr=L
  \output={\if L\lr
        \global\setbox\leftcolumn=\columnbox\global\let\lr=R \advancepageno
      \else \doubleformat \global\let\lr=L\fi
    \ifnum\outputpenalty>-20000 \else\dosupereject\fi}
  \def\doubleformat{\shipout\vbox{
    \fullline{\box\leftcolumn\hfil\columnbox}\advancepageno}}
  \def\columnbox{\leftline{\vbox{\makeheadline\pagebody\makefootline}}}
  \tolerance=1000 }
\twelvepoint
\doublespace
{\nopagenumbers{
\rightline{IASSNS-HEP-00/62}
\rightline{~~~September, 2000}
\bigskip\bigskip
\centerline{\rmb Completing the Square to Find the Supersymmetric 
Matter Effective} 
\centerline{\rmb Action Induced by Coupling to Linearized $N=1$ 
Supergravity}

\medskip
\centerline{\it  Stephen L. Adler
}
\centerline{\bf Institute for Advanced Study}
\centerline{\bf Princeton, NJ 08540}
\medskip
\leftline{{\it Short title:} Completing the Square}
\bigskip\bigskip
\leftline{\it Send correspondence to:}
\medskip
{\singlespace\leftline{Stephen L. Adler}
\leftline{Institute for Advanced Study}
\leftline{Einstein Drive, Princeton, NJ 08540}
\leftline{Phone 609-734-8051; FAX 609-924-8399; 
email adler@ias.edu}}
\bigskip\bigskip
}}
\vfill\eject
\pageno=2
\AB
We consider generic $N=1$ supersymmetric matter coupled to linearized 
$N=1$ supergravity through the multiplet of currents.    
By completing the square, we find the effective action giving the leading 
supergravity induced correction to the matter dynamics, expressed explicitly 
as a quadratic form in the components of the current multiplet.   
The effective action is supersymmetry invariant through an 
interplay of the local terms arising from the auxiliary field couplings, 
and the 
nonlocal terms arising from graviton and gravitino exchange, neither   
of which is separately invariant.  Having an explicit form for the 
supergravity induced effective action is a first step in studying 
whether supergravity corrections can lead to dynamical supersymmetry 
breaking in supersymmetric matter dynamics.  In Appendices we give 
explicit expressions for the currents, in our notational conventions, in the 
Wess-Zumino and supersymmetric Yang Mills models.  
\AE
\bigskip\bigskip
\vfill\eject
\pageno=3
\centerline{{\bf 1.~~Introduction}}
Supersymmetry, if it is to be relevant to physics, must be broken, and 
mechanisms for supersymmetry breaking have been a subject of ongoing study.  
While the familiar O'Raifeartaigh and Fayet-Iliopoulos 
mechanisms [1] rely on the presence of scalar 
components of chiral matter supermultiplets, or $U(1)$ gauge supermultiplet 
auxiliary fields, respectively, there still exists the possibility that 
supersymmetry may be dynamically broken in theories lacking these fields.  
In particular, for supersymmetric non-Abelian gauge theories, while there  
are general index theorem arguments which show that such theories 
cannot break supersymmetry dynamically when considered in isolation [1], 
such theorems do not apply to supersymmetric non-Abelian theories coupled  
to supergravity.  Thus, there remains the interesting possibility that 
supergravity couplings may trigger dynamical symmetry breaking in 
supersymmetric gauge theories.  

The aim of this paper is to carry out the first technical step needed in 
a study of whether supergravity couplings can induce matter 
supersymmetry breaking,  by integrating out the supergravity dynamics to 
leading order to give a supersymmetric effective matter action, which 
incorporates the effects of the supergravity couplings.  
Although calculating the supergravity 
induced effective action involves no fundamentally new concepts, it  
appears not to have been done before. Since the results are elegant, 
and illustrate the intimate connection between the supercurrent multiplet 
and the linearized supergravity multiplet including auxiliary fields, 
we present a detailed account of them here.    We carry out our 
calculations in terms of component fields, with careful attention to 
such issues as the phases appearing in the supercurrent transformation,  
the effect of gauge invariances of linearized supergravity on the 
inversions of the kinetic terms needed to isolate the effective action, and 
the independence of the results of the choice of gauge fixing.  
In future work, we plan to study the effective action derived in this paper,  
to see  whether  
it permits evasion of the classical ``no-go'' theorems restricting 
dynamical symmetry breaking in supersymmetric theories.  Well-known   
examples in non-supersymmetric theories where analogous effective actions 
lead to symmetry breaking are the BCS theory of superconductivity, where the 
phonon exchange effective action is responsible for symmetry breaking, and 
models for chiral symmetry breaking, dynamical electroweak symmetry breaking, 
and ``color superconductivity'' in non-Abelian gauge theories, where the 
one gluon exchange effective action leads to symmetry breaking.  

In Sec. 2 we review the transformation properties of the linearized 
supergravity multiplet formulated using the minimal auxiliary fields, 
as well as the analogous transformation properties of 
the supercurrent multiplet, and give the standard interaction 
Lagrangian which is invariant under 
simultaneous supersymmetry transformations of the supergravity and 
supercurrent multiplets.  We also summarize the Abelian gauge invariances 
which incorporate the effect of general coordinate invariance in the 
linearized theory, and the closely related conservation laws for the 
supercurrent multiplet,  which render the interaction Lagrangian gauge 
invariant.  In Sec. 3 we consider, as an analog, the familiar case of 
quantum electrodynamics, and show that current conservation permits 
one to complete the square to find the effective action giving the 
effects of the photon dynamics on the charged fermions, independently of 
the choice of gauge fixing action.  In Sec. 4 we show how 
to carry out the analogous calculation   
in the linearized supergravity case, by using energy momentum tensor and 
supersymmetry current conservation to complete the square in the 
graviton and gravitino kinetic terms, independently of the 
choice of the supergravity  gauge fixing action.  
In Sec. 5 we summarize the resulting 
formula for the effective action, which is explicitly invariant 
under the supersymmetry transformation of the supercurrent multiplet.  
In Appendix A we give our metric and gamma matrix conventions,   
in Appendix B we give formulas for the supercurrent components and the 
related ``supercurrent anomalies''  in the Wess-Zumino model, and in 
Appendix C we give analogous formulas for the supersymmetric Yang-Mills 
model. These currents obey the transformation formulas given in Sec. 2 
(and were used as a check on the phases appearing in these transformation 
formulas), and will be needed in applications of the effective action to 
the study of the possibility of supergravity induced dynamical symmetry 
breaking in the respective models to which they pertain.  

\bigskip
\centerline{\bf 2.~~The linearized supergravity multiplet and 
the supercurrent multiplet}
In linearized general relativity, the spacetime metric $g_{\mu\nu}$ 
is assumed to deviate from the Minkowski metric $\eta_{\mu\nu}$
by only a small perturbation proportional to $h_{\mu\nu}$ , 
$$g_{\mu\nu}=\eta_{\mu\nu}+2\kappa h_{\mu\nu}~~~.\eqno(2.1)$$
The constant $\kappa$ appearing in Eq.~(2.1) is defined by  
$$\kappa=(8\pi G)^{1\over 2}=M_{\rm Planck}^{-1}~~~,\eqno(2.2)$$
where $G$ is Newton's constant, and so the perturbation $h_{\mu\nu}$ 
has dimension one, as is usual for a bosonic field.  In linearized 
supergravity, one adjoins to the spin 2 graviton field $h_{\mu\nu}$ 
a spin $3/2$ Rarita-Schwinger Majorana field $\psi_{\mu}$, which 
describes the 
fermionic gravitino partner of the graviton.  Although one can write 
down a supersymmetry algebra based on just the graviton and gravitino 
fields, it closes only ``on shell'', that is, with use of the equations 
of motion.  

To get a supergravity multiplet for which the supersymmetry 
algebra closes without use of the equations of motion, it is necessary 
to add auxiliary fields which vanish on shell in the absence of matter  
source couplings.  The minimal set of auxiliary fields for supergravity 
has been shown [2] to be an axial vector $b_{\mu}$, a scalar $M$, and a 
pseudoscalar $N$; the supersymmetry variations of these fields and 
of $h_{\mu\nu}$ and $\psi_{\mu}$, which represent the supersymmetry 
algebra without invoking the equations of motion, are
$$\eqalign{
\delta h_{\mu\nu}=&{1\over 2}\overline{\epsilon}(\gamma_{\mu}\psi_{\nu}
+\gamma_{\nu}\psi_{\mu})~~~,\cr
\delta \psi_{\mu}=&[-\sigma^{\kappa\nu}\partial_{\kappa}h_{\nu\mu}
-{1\over 3}\gamma_{\mu}(M+i\gamma_5 N)+(b_{\mu}-{1\over 3} \gamma_{\mu}
\gamma\cdot b) i \gamma_5] \epsilon~~~,\cr
\delta \overline{\psi}_{\mu}=&\overline{\epsilon}[\sigma^{\kappa\nu}
\partial_{\kappa}h_{\nu\mu}+{1\over 3}(M+i\gamma_5N)\gamma_{\mu} 
+i\gamma_5(b_{\mu}
-{1\over 3}\gamma \cdot b \gamma_{\mu})]~~~,\cr
\delta b_{\mu}=&{3 \over 2}i\overline{\epsilon}\gamma_5(R_{\mu}-
{1\over 3} \gamma_{\mu} \gamma \cdot R)~~~,\cr
\delta M=&-{1\over 2} \overline \epsilon \gamma \cdot R~~~,\cr
\delta N=&-{1\over 2}i \overline{\epsilon} \gamma_5 \gamma \cdot R~~~.\cr
}\eqno(2.3)$$
Here $\epsilon$ is a constant Grassmann supersymmetry parameter (in 
linearized supergravity, the supersymmetry transformations represent a 
global supersymmetry), the $\gamma_{\mu}$ are the usual Dirac matrices 
and $\sigma_{\mu\nu}$ are proportional to their commutators (see Appendix 
A for details of our conventions), and $R_{\mu}$ is defined by 
$$R^{\nu}=i\epsilon^{\nu\mu\kappa\rho}\gamma_5\gamma_{\mu}\partial_{\kappa}
\psi_{\rho}~~~.\eqno(2.4)$$
A straightforward calculation [1] now shows that the 
transformations of Eq.~(2.3) 
are an invariance of the linearized supergravitational action 
$$S_{\rm grav}=\int d^4x[E^{\mu\nu}h_{\mu\nu}-{1\over 2}\overline{\psi}_{\mu}
R^{\mu}-{1\over 3}(M^2+N^2-b_{\mu}b^{\mu})]~~~,\eqno(2.5)$$
with $E^{\mu\nu}$ the linearized Einstein tensor defined by 
$$E_{\mu\nu}={1\over 2}(\partial_{\mu}\partial_{\nu}h_{\lambda}^{\lambda}
+\square h_{\mu\nu}-\partial_{\mu}\partial^{\lambda}h_{\lambda\nu}
-\partial_{\nu}\partial^{\lambda}h_{\lambda\mu}-\eta_{\mu\nu}\square 
h_{\lambda}^{\lambda}+\eta_{\mu\nu}\partial^{\lambda}\partial^{\rho}
h_{\lambda\rho})~~~.\eqno(2.6)$$

Linearized supergravity couples to supersymmetric matter through a real
supermultiplet of currents [3, 4], consisting of the energy momentum tensor 
$\theta^{\mu\nu}$, the supersymmetry current $j_{\mu}$, an axial 
vector current (the $R$ symmetry current) $j_{\mu}^{(5)}$, a scalar 
density $P$ and a pseudoscalar density $Q$.  These transform under 
supersymmetry variations as 
$$\eqalign{
\delta \theta^{\mu\nu}=&{1\over 4}\overline{\epsilon}
(\sigma^{\kappa\mu}\partial_{\kappa}j^{\nu}+\sigma^{\kappa\nu}
\partial_{\kappa}j^{\mu})~~~,\cr
\delta j_{\mu}=&[2\gamma^{\nu}\theta_{\mu\nu}-i\gamma_5\gamma\cdot\partial
j_{\mu}^{(5)}+i\gamma_5\gamma_{\mu}\partial \cdot j^{(5)}
+{1\over 2}\epsilon_{\mu\nu\rho\kappa}\gamma^{\nu}\partial^{\rho}
j^{\kappa (5)} +{1 \over 3}\sigma_{\mu\nu}\partial^{\nu}(P+i\gamma_5 Q)]
\epsilon~~~,\cr
\delta \overline{j}_{\mu}=&\overline{\epsilon}[-2\theta_{\mu\nu}\gamma^{\nu}
-i\gamma_5 \gamma \cdot \partial j_{\mu}^{(5)} + i \gamma_5\gamma_{\mu}
\partial \cdot j^{(5)}-{1\over 2}\epsilon_{\mu\nu\rho\kappa}\gamma^{\nu}
\partial^{\rho}j^{\kappa(5)}-{1\over 3} \sigma_{\mu\nu}\partial^{\nu}
(P+i\gamma_5 Q)] ~~~,\cr
\delta j_{\mu}^{(5)}=&i\overline{\epsilon}\gamma_5 j_{\mu}
-{1\over 3}i \overline{\epsilon}\gamma_5 \gamma_{\mu} \gamma \cdot j~~~,\cr
\delta P=&\overline{\epsilon} \gamma \cdot j~~~,\cr
\delta Q=&i\overline{\epsilon}\gamma_5 \gamma \cdot j~~~.
}\eqno(2.7)$$
Calculating from Eq.~(2.7), one finds that the  ``anomaly'' chiral 
supermultiplet
consisting of $\theta_{\mu}^{\mu}$, $\gamma \cdot j$, 
$\partial \cdot j^{(5)}$, $P$, and $Q$, has the supersymmetry variations 
$$\eqalign{
\delta \theta_{\mu}^{\mu}=&{1\over 2} \overline {\epsilon} 
\gamma \cdot \partial \gamma \cdot j~~~,\cr
\delta \gamma \cdot j=&[2 \theta_{\mu}^{\mu}-3i\gamma_5 
\partial \cdot j^{(5)}
+\gamma \cdot \partial (P+i\gamma_5 Q)]\epsilon~~~,\cr
\delta \overline{j}\cdot \gamma=&\overline{\epsilon}[-2 \theta_{\mu}^{\mu}
+3i\gamma_5 \partial \cdot j^{(5)}+\gamma \cdot 
\partial(P-i\gamma_5 Q)]~~~,\cr
\delta \partial \cdot j^{(5)}=&-{1\over 3}i \overline{\epsilon} \gamma_5
\gamma \cdot \partial \gamma \cdot j~~~,\cr
\delta P=&\overline{\epsilon} \gamma \cdot j~~~,\cr
\delta Q=&i\overline{\epsilon}\gamma_5 \gamma \cdot j~~~.
}\eqno(2.8)$$

A direct calculation now verifies that the matter interaction action that 
is invariant under the simultaneous supersymmetry variations of 
Eqs.~(2.3) and (2.7) is 
$$S_{\rm int}=\kappa \int d^4x  [h_{\mu\nu}\theta^{\mu\nu}
+{1 \over 2} \overline{\psi}_{\mu}j^{\mu}-{1\over 2} b_{\mu}j^{\mu(5)}
-{1\over 6}(MP+NQ)]~~~.\eqno(2.9)$$
The overall normalization of Eq.~(2.9) is determined by the requirement 
that Eqs.~(2.5) and (2.9) give the correct Newtonian potential in the 
static limit; this will be verified explicitly in Sec. 4.  Since the 
auxiliary fields $b_{\mu}$, $M$, and $N$ enter Eqs.~(2.5) and (2.9) 
with no differential operators acting on them, their equations of motion 
are simply the algebraic relations 
$$M=-{1\over 4} \kappa P~,~~N=-{1\over 4} \kappa Q~,~~b_{\mu}={3\over 4}
\kappa j_{\mu}^{(5)}~~~.\eqno(2.10)$$
Using these relations to eliminate the auxiliary fields, we get
$$\eqalign{
S_{\rm tot}=&S_{\rm grav}+S_{\rm int} \cr
=&\int d^4x [E^{\mu\nu}h_{\mu\nu}-{1\over 2}\overline{\psi}_{\mu}R^{\mu}
+\kappa (h_{\mu\nu}\theta^{\mu\nu} 
+{1\over 2} \overline {\psi}_{\mu} j^{\mu})
-{3\over 16} \kappa^2 j_{\mu}^{(5)}j^{\mu(5)}
+{1\over 48} \kappa^2(P^2+Q^2)]~~~.\cr
}\eqno(2.11)$$
Our aim in this paper will be to integrate out the dynamical graviton 
and gravitino fields, thus transforming 
Eq.~(2.11) into the complete order $\kappa^2$ effective action giving the 
effect of supergravity couplings on the supersymmetric matter fields.  

The reason this integration is nontrivial is that the actions of Eqs.~(2.5),  
(2.9), and (2.11) have two Abelian gauge invariances, which 
are the reflection in 
the linearized theory of general coordinate invariance [1].  
These invariances 
are 
$$\eqalign{
h_{\mu\nu} \to& h_{\mu\nu} + \partial _{\mu} \Phi_{\nu} + \partial _{\nu}
\Phi_{\mu}~~~,\cr
\psi_{\mu} \to&\psi_{\mu}+\partial_{\mu} \Psi~~~,\cr
}\eqno(2.12)$$
with $\Phi_{\nu}$ and $\Psi$ arbitrary real four vector and Majorana spinor 
gauge functions, respectively.  The invariance 
of $S_{\rm grav}$ follows directly from  Eqs.~(2.4)--(2.6) and 
(2.12), while the invariance of $S_{\rm int}$ is a consequence of 
Eqs.~(2.9) and (2.12), together with the energy momentum tensor and the 
supersymmetry current conservation relations 
$$\partial_{\mu}\theta^{\mu\nu}=\partial_{\nu}\theta^{\mu\nu}
=\partial_{\mu}j^{\mu}=0~~~.\eqno(2.13)$$ 
In order to integrate out the graviton and gravitino fields to obtain the 
complete effective action, we will have 
to take the existence of these Abelian gauge invariances into account.

\bigskip
\centerline{\bf 3.~~Quantum electrodynamics as an instructive analog}

Since the Abelian gauge invariances of Eq.~(2.12) resemble the 
familiar gauge invariance of quantum electrodynamics (QED), it is instructive 
to consider the calculation of the effective action in QED as an example.  
We start from the action 
$$S_{\rm QED}=\int d^4x(A^{\alpha}{1\over 2} P_{\alpha \beta} A^{\beta}
+j_{\alpha}A^{\alpha})~~~,\eqno(3.1)$$
with $A^{\alpha}$ the Abelian gauge potential, with $P_{\alpha\beta}$ 
the differential operator 
$$P_{\alpha \beta}=\square \eta_{\alpha\beta}
-\partial_{\alpha}\partial_{\beta}~~~,\eqno(3.2)$$
and with $j_{\alpha}$ a conserved source current, 
$$\partial^{\alpha}j_{\alpha}=0~~~.\eqno(3.3)$$
By virtue of the structure of $P_{\alpha\beta}$ and the conservation of 
the source current $j_{\alpha}$, the action of Eq.~(3.1) is invariant 
under the Abelian gauge transformation
$$A_{\alpha} \to A_{\alpha}+\partial_{\alpha} \Phi~~~,\eqno(3.4)$$
with $\Phi$ an arbitrary real gauge function.   As a result of this gauge 
invariance, $P_{\alpha\beta}$ is not an invertible operator, necessitating 
the inclusion of a gauge fixing term  when the action of Eq.~(3.1) is used 
in a Feynman path integral.  However, because the current $j_{\alpha}$ is 
conserved, we can nonetheless complete the square in the kinetic term of 
the action by writing 
$$\eqalign{
A^{\alpha}{1\over 2}P_{\alpha\beta} A^{\beta} +j_{\alpha}A^{\alpha} 
=&{1\over 2}(A^{\alpha}+j^{\alpha}{1\over \square}) P_{\alpha\beta}
(A^{\beta}+{1\over \square} j^{\beta})-{1\over 2}j^{\alpha}{1\over \square}
j_{\alpha}~~~.\cr
}\eqno(3.5)$$

When Eqs.~(3.1) and (3.5) are inserted in a functional 
integral, one can use the 
translation invariance of the functional measure to define a new integration 
variable 
$$A^{\prime \alpha}\equiv A^{\alpha} +{1 \over \square} j^{\alpha}~~~,
\eqno(3.6)$$
and so the $P_{\alpha \beta}$ term on the right hand side of 
Eq.~(3.5) contributes a $j^{\alpha}$ independent 
constant factor to the functional integral.  This argument is independent 
of the choice of gauge fixing action, since the Faddeev-Popov gauge fixing 
procedure is simply a method for isolating the integral 
over the gauge orbit [5], which in the Abelian case is unchanged 
by a constant 
translation.  In the special case in which one employs the ``standard'' 
covariant gauge fixing action 
$$S_{\rm fix}={1\over 2}\alpha_s (\partial \cdot A)^2~~~,\eqno(3.7)$$
the gauge fixing action itself is unchanged in form by the substitution of 
Eq.~(3.6), independent of the value of the gauge parameter $\alpha_s$, 
because of current conservation.  For generic gauge fixings, the $P_{\alpha 
\beta}$ term on the right hand side of Eq.~(3.5) makes a 
contribution that is independent of the source current 
only after the functional integral has been carried out.    

Thus, by completing the square and making the change of 
variable of Eq.~(3.6),  we have learned that, after doing the 
gauge field functional integral, the source current dependence is 
contained solely in the second term of Eq.~(3.5),  giving 
$$S_{\rm eff}=-\int d^4x {1\over 2} j^{\alpha} {
1\over \square} j_{\alpha}~~~.
\eqno(3.8)$$ 
Explicitly indicating the space time arguments, this takes the form 
$$S_{\rm eff}=\int d^4x d^4y {1\over 2} j^{\alpha}(x)\Delta_F(x-y)j_{\alpha}
(y)~~~,\eqno(3.9)$$
where we have introduced the Feynman propagator $\Delta_F(x-y)$ given by 
$$\Delta_F(x-y)={1\over (2\pi)^4}\int d^4q {e^{i q \cdot (x-y)}
\over q^2-i0^+}~~~,\eqno(3.10)$$
which obeys 
$$\square \Delta_F(x-y)=-\delta^4(x-y)~~~.\eqno(3.11)$$
(The use of the Feynman $i0^+$ contour prescription in inverting $\square$ 
is dictated, as usual, by the requirements of relativistic invariance and 
causality.)  We note that Eq.~(3.9) is just the result that would be 
obtained from the usual covariant Feynman rules, since the propagator 
corresponding to the standard gauge fixing of Eq.~(3.7) is proportional to   
$(\delta_{\alpha\beta}+...)\Delta_F$, with $...$ indicating $\alpha_s$ 
dependent derivative terms that vanish when acting on the conserved source 
currents.  

As a check on normalizations, let us verify that Eq.~(3.9) leads to the 
correct Coulomb force law in the static limit when 
$j^0(x)=-j_0(x)=\rho(\vec x)$,  $\vec j=0$.  Writing 
$$\int dx^0dy^0 = \int dt d(x^0-y^0)~~~,\eqno(3.12)$$
with $t={1\over 2}(x^0+y^0)$, and using 
$$\int d(x^0-y^0) \Delta_F(x-y)={1 \over 4 \pi |\vec x -\vec y|}~~~,
\eqno(3.13)$$
the static limit of Eq.~(3.9) becomes 
$$S_{\rm eff}=\int dt \int d^3x d^3y (-{1\over 2}) 
{\rho(\vec x) \rho(\vec y) \over 4 \pi |\vec x -\vec y|}~~~.\eqno(3.14)$$
This agrees in both sign and magnitude with the action contribution 
$S=-\int dt V_{\rm Coulomb}$ arising from a static charge distribution 
$\rho(\vec x)$. 

We are now ready to return to the gravitational case, where we will see that 
every aspect of the familiar QED calculation that we have just 
reviewed has a direct analog.

\bigskip
\centerline{\bf 4.~~Completing the square for the graviton and gravitino}

We now apply the lessons learned in the preceding section to the  
gravitino and graviton terms [6] in Eq.~(2.11).  
Beginning with the gravitino, 
let us introduce the invertible kernel $M^{\nu\kappa}$ defined by 
$$\eqalign{
M^{\nu\kappa}=&-(\gamma^{\nu}\partial^{\kappa}+\gamma^{\kappa}
\partial^{\nu})+\eta^{\nu\kappa}\gamma \cdot \partial 
+ {1\over 2}\gamma^{\nu}
\gamma \cdot \partial \gamma^{\kappa} \cr
=&-{1\over 2} \gamma^{\kappa} \gamma \cdot \partial \gamma^{\nu}~~~,\cr
}\eqno(4.1)$$
which obeys
$$M_{\mu\nu}M^{\nu\kappa}=\square \delta_{\mu}^{\kappa}~~~.\eqno(4.2)$$
In terms of $M_{\mu\nu}$, we define a second kernel $N_{\mu\nu}$ by 
$$N_{\mu\nu}=-{1\over 2}i \epsilon_{\mu\nu}^{~~~\eta\rho}\partial_{\eta}
\gamma_{\rho}\gamma_5=-{1\over 2}\left(M_{\mu\nu}+{1\over 2} \gamma_{\mu}
\gamma \cdot \partial \gamma_{\nu}\right)~~~,\eqno(4.3)$$
which obeys 
$$N_{\mu\nu}M^{\nu\theta}=-{1\over 2} \square \delta_{\mu}^{~\theta}
+{1 \over 2}\gamma_{\mu} \gamma \cdot \partial \partial^{\theta}~~~.
\eqno(4.4)$$
Since the kinetic term for the gravitino in Eq.~(2.11) can be expressed 
it terms of the kernel $N_{\mu\nu}$, 
$$\int d^4x (-{1\over 2}\overline{\psi}_{\mu}R^{\mu})=\int d^4x  
\overline{\psi}^{\mu} N_{\mu\nu} \psi^{\nu}~~~,\eqno(4.5)$$
and since the second term on the right of Eq.~(4.4) contains a factor 
$\partial^{\theta}$ that gives zero when acting on the conserved 
supersymmetry current  $j_{\theta}$, we can use Eq.~(4.4) to complete 
the square for the gravitino. Using Eq.~(A.7) to relate the adjoint of  
$M^{\mu\tau}$ to $M^{\tau\mu}$, we get 
$$\eqalign{
\overline{\psi}^{\mu} N_{\mu\nu} \psi^{\nu}
+{1\over 2}\kappa \overline{\psi}_{\mu}
j^{\mu}
=&\overline{\left(\psi^{\mu}-{1\over 2} \kappa M^{\mu\tau} {1\over \square}
j_{\tau}\right)} N_{\mu\nu}\left(
\psi^{\nu}-{1\over 2} \kappa M^{\nu\theta} 
{1 \over \square} j_{\theta}
\right) \cr  
+&{1\over 8} \kappa^2 \overline{j}_{\tau} M^{\tau\nu} {1 \over \square} 
j_{\nu}~~~.\cr
}\eqno(4.6)$$

When Eqs.~(2.11), (4.5), and (4.6) are inserted in a functional integral, 
we can use translation invariance of the functional measure to define 
a new integration variable 
$$\psi^{\prime \nu}=
\psi^{\nu}-{1\over 2} \kappa M^{\nu\theta} 
{1 \over \square} j_{\theta}~~~,\eqno(4.7)$$
and so the $N_{\mu\nu}$ term on the right hand side of 
Eq.~(4.6) contributes a $j_{\nu}$ 
independent constant factor to the functional integral.  As in the QED   
case this argument is independent of the choice of gauge fixing.  In the 
special case in which one employs the ``standard'' Rarita-Schwinger gauge 
fixing action 
$$S_{\rm fix}={1\over 2} \alpha_{\rm RS}\int d^4x \overline{\psi}^{\mu}
{1 \over 2} \gamma_{\mu} \gamma \cdot \partial \gamma_{\nu} \psi^{\nu}~~~,
\eqno(4.8)$$
the fact that 
$$ \gamma_{\nu}M^{\nu \theta} \propto  \partial^{\theta}
~~~\eqno(4.9)$$
implies that by current conservation, 
the gauge fixing action of Eq.~(4.8) is unchanged 
in form by the change of variable of Eq.~(4.7).  

We conclude that by completing the square and making the change of 
variable of Eq.~(4.7), after doing the gravitino functional integral the 
supersymmetry current dependence is contained solely in the second term 
on the right hand side of Eq.~(4.6), giving 
$$\eqalign{
S_{\rm eff}=&
{1\over 8} \kappa^2 \overline{j}_{\tau} M^{\tau\nu} {1 \over \square} 
j_{\nu}\cr
=&-{1\over 8} \kappa^2 \int d^4x d^4y \overline{j}_{\tau}(x) 
\left(\eta^{\tau\nu} \gamma \cdot \partial_{x}+{1\over 2} \gamma^{\tau} 
\gamma \cdot \partial_{x} \gamma^{\nu}\right) \Delta_F(x-y) j_{\nu}(y)
~~~,\cr
}\eqno(4.10)$$
where we have dropped terms that vanish by conservation of the 
supersymmetry current $j_{\nu}$.  This is just the result that would be 
obtained by the usual covariant Feynman rules, since the gravitino 
propagator corresponding to the gauge fixing of Eq.~(4.8) is proportional 
to $M^{\tau \nu} \Delta_F+...$, with $...$ indicating $\alpha_{\rm RS}$ 
dependent derivative terms that vanish when acting on the 
conserved supersymmetry currents.  

We turn next to the analogous completion of square argument for the 
graviton.  We rewrite the graviton kinetic term of Eq.~(2.11) in the form 
$$\int d^4x E^{\mu\nu}h_{\mu\nu}=\int d^4x h^{\alpha\beta} 
P_{\alpha\beta,\mu\nu}h^{\mu\nu}~~~,\eqno(4.11)$$
with $P_{\alpha\beta,\mu\nu}$ the differential operator 
$$\eqalign{
P_{\alpha\beta,\mu\nu}=&{1\over 2}(\eta_{\alpha \beta} \partial_{\mu}
\partial_{\nu}+\eta_{\mu\nu}\partial_{\alpha}\partial_{\beta})
+{1\over 4} \square(\eta_{\alpha\mu} \eta_{\beta \nu}+ \eta_{\alpha \nu}
\eta_{\beta \mu})  \cr
-&{1\over 4}(\partial_{\mu}\partial_{\alpha} \eta_{\beta \nu}
+\partial_{\nu}\partial_{\alpha} \eta_{\beta \mu}
+\partial_{\mu}\partial_{\beta}\eta_{\alpha\nu}
+\partial_{\nu}\partial_{\beta} \eta_{\alpha\mu})
-{1\over 2} \eta_{\mu\nu}\eta_{\alpha\beta} \square~~~.\cr
}\eqno(4.12)$$
Introducing now the projector $Q^{\mu\nu,\gamma\delta}$ defined by 
$$Q^{\mu\nu,\gamma\delta}=\eta^{\mu\gamma}\eta^{\nu\delta} + 
\eta^{\mu\delta}\eta^{\nu\gamma}-\eta^{\mu\nu}\eta^{\gamma\delta}~~~,
\eqno(4.13)$$
we find that 
$$P_{\alpha\beta,\mu\nu}Q^{\mu\nu,\gamma\delta}=
{1\over 2}\square (\delta_{\alpha}^{\gamma}\delta_{\beta}^{\delta}
+\delta_{\beta}^{\gamma}\delta_{\alpha}^{\delta})
+{1\over 2}z_{\alpha\beta}^{~~~\delta}\partial^{\gamma}
+{1\over 2}z_{\alpha\beta}^{~~~\gamma}\partial^{\delta}~~~,\eqno(4.14)$$
with 
$$z^{\alpha\beta\delta}=\eta^{\alpha\beta}\partial^{\delta}
-(\eta^{\delta\beta}\partial^{\alpha}+\eta^{\delta\alpha}\partial^{\beta})
~~~.\eqno(4.15)$$
Since the $z_{\alpha\beta}^{~~~\gamma,\delta}$ terms in Eq.~(4.14) 
contain derivatives $\partial^{\gamma}~,~
\partial^{\delta}$ that give zero when acting on the conserved energy 
momentum tensor $\theta_{\gamma \delta}$, we can complete the square 
for the graviton as follows, 
$$\eqalign{
&h^{\alpha\beta}P_{\alpha\beta,\mu\nu}h^{\mu\nu} + 
\kappa  h_{\alpha\beta}  \theta^{\alpha\beta} \cr
=&\left( h^{\alpha\beta}+\kappa \theta_{\nu\tau} {1 \over 2\square}
Q^{\nu\tau,\alpha\beta}\right) P_{\alpha\beta,\mu\nu}
\left(h^{\mu\nu}+Q^{\mu\nu,\gamma\delta} {1\over 2\square} \kappa 
\theta_{\gamma\delta}\right)\cr
-&{1\over 4} \kappa^2 \theta^{\nu\tau} {1 \over \square} Q_{\nu\tau,\alpha
\beta}\theta^{\alpha\beta}~~~.\cr
}\eqno(4.16)$$

Again, when Eqs. (2.11), (4.11), and (4.16) are inserted in a functional 
integral, one can use translation invariance of the functional measure  
to define a new integration variable 
$$h^{\prime \mu\nu}=
h^{\mu\nu}+Q^{\mu\nu,\gamma\delta} {1\over 2\square} \kappa 
\theta_{\gamma\delta}~~~,\eqno(4.17)$$
and so the $P_{\alpha\beta,\mu\nu}$ term on the right hand side of 
Eq.~(4.16) contributes an 
energy momentum tensor independent constant factor to the functional 
integral.  As before, this argument is independent of the choice of 
gauge fixing action.  Again, in the special case that one makes the 
``standard'' gauge fixing choice\footnote{*}{We remark that there is no  
linear combination of the gauge fixing actions of Eq.~(4.8) 
and Eq.~(4.18) which 
is supersymmetric, even ``on shell'' when the equations of motion are used.}
$$S_{\rm fix} = {\alpha_{\rm G} \over \kappa^2} 
\int d^4x \eta_{\mu\nu}\partial_{\alpha}(\surd g g^{\alpha\mu}) 
\partial_{\beta}(\surd g g^{\beta \nu})~~~,\eqno(4.18)$$
which after linearization is proportional to 
$$\eqalign{
&{1\over 2}\alpha_{\rm G}\int d^4x[(\partial_{\alpha}h^{\alpha\mu})^2
+h_{\theta}^{\theta}\partial_{\alpha}\partial_{\mu}h^{\alpha\mu}
+{1\over 4}(\partial^{\mu}h_{\theta}^{\theta})^2] \cr
=&{1\over 2} \alpha_{\rm G} \int d^4x h^{\alpha\beta} G_{\alpha\beta,\mu\nu}
h^{\mu\nu}~~~,\cr
}\eqno(4.19)$$
the gauge fixing action is unchanged in form by the change of variables 
of Eq.~(4.17).  This follows from the fact that the projector introduced 
in Eq.~(4.19), 
$$\eqalign{
G_{\alpha\beta,\mu\nu}=&{1\over 2}
(\eta_{\alpha \beta} \partial_{\mu}\partial_{\nu}
+\eta_{\mu\nu}\partial_{\alpha}\partial_{\beta})
-{1\over 4} \eta_{\mu\nu}\eta_{\alpha\beta} \square \cr  
-&{1\over 4}(\partial_{\mu}\partial_{\alpha} \eta_{\beta \nu}
+\partial_{\nu}\partial_{\alpha} \eta_{\beta \mu}
+\partial_{\mu}\partial_{\beta}\eta_{\alpha\nu}
+\partial_{\nu}\partial_{\beta} \eta_{\alpha\mu})\cr
=&P_{\alpha\beta,\mu\nu}-{1\over 4} \square Q_{\alpha\beta,\mu\nu}~~~,\cr
}\eqno(4.20)$$
obeys
$$G_{\alpha\beta,\mu\nu}Q^{\mu\nu,\gamma\delta} \propto 
z_{\alpha\beta}^{~~~\delta}\partial^{\gamma}
+z_{\alpha\beta}^{~~~\gamma}\partial^{\delta}
~~~,\eqno(4.21)$$ 
and hence vanishes when acting on the conserved energy momentum tensor. 

We again conclude that by completing the square and making the change of 
variable of Eq.~(4.17), after doing the graviton functional integral the 
energy momentum tensor dependence is contained solely in the second term 
on the right hand side of Eq.~(4.16), giving 
$$\eqalign{
S_{\rm eff}=&-{1\over 4} \kappa^2 \theta^{\nu\tau} 
{1\over \square} Q_{\nu\tau,\alpha\beta} \theta^{\alpha\beta} \cr
=&{1\over 4} \kappa^2 \int d^4x d^4y 
\theta^{\nu\tau}(x) 
(\eta_{\nu\alpha}\eta_{\tau\beta}+ \eta_{\nu\beta}\eta_{\tau\alpha}
-\eta_{\nu\tau}\eta_{\alpha\beta})\Delta_F(x-y) 
\theta^{\alpha\beta}(y)~~~.\cr
}\eqno(4.22)$$
This is again the result that would be obtained by using the standard 
covariant Feynman rules,  
since the graviton  
propagator corresponding to the gauge fixing of Eq.~(4.18) is proportional 
to $Q_{\nu\tau,\alpha\beta}\Delta_F+...$, with $...$ indicating 
$\alpha_{\rm G}$ 
dependent derivative terms that vanish when acting on the conserved energy   
momentum tensor.  

As a check on the normalization, let us verify that the static limit of 
Eq.~(4.22) agrees with the Newtonian potential.  Considering the case 
when $\theta_{00}=\rho(\vec x)$, $\theta_{0j}=\theta_{ij}=0$, and using 
$Q^{00,00}=1$ and $\kappa^2/(8\pi)=G$ together with Eqs.~(3.12)-(3.13), 
we get 
$$S_{\rm eff}=\int dt \int d^3xd^3y {1\over 2} G 
{\rho(\vec x) \rho(\vec y) \over  |\vec x -\vec y|}~~~,\eqno(4.23)$$
agreeing with the action contribution $S=-\int dt V_{\rm Newton}$ 
arising from a static mass distribution $\rho(\vec x)$.  

\bigskip
\centerline{\bf 5.~~The effective action and its supersymmetry invariance}
We are now ready to assemble our final result.  Combining the order 
$\kappa^2$ terms in Eq.~(2.11) with the gravitino and graviton effective 
actions of Eqs.~(4.10) and 4.22), we get for the complete order $\kappa^2$ 
effective action the result 
$$\eqalign{
\kappa^{-2}S_{\rm eff}=&\int d^4x \left[-{3\over 16} j_{\mu}^{(5)}j^{\mu(5)}
+{1\over 48} (P^2+Q^2)\right]\cr
+&\int d^4x d^4y \left[ {1\over 4}\theta^{\nu\tau}(x) 
(\eta_{\nu\alpha}\eta_{\tau\beta}+ \eta_{\nu\beta}\eta_{\tau\alpha}
-\eta_{\nu\tau}\eta_{\alpha\beta})\Delta_F(x-y) 
\theta^{\alpha\beta}(y)\right.\cr
-&\left.{1\over 8}  \overline{j}_{\tau}(x) 
\left(\eta^{\tau\nu} \gamma \cdot \partial_{x}+{1\over 2} \gamma^{\tau} 
\gamma \cdot \partial_{x} \gamma^{\nu}\right) \Delta_F(x-y) j_{\nu}(y)
\right]~~~.\cr
}\eqno(5.1)$$
By use of the current multiplet supersymmetry transformation given in 
Eq.~(2.7), it is straightforward to verify that the effective 
action of Eq.~(5.1) is supersymmetry invariant when the conservation 
relations of Eq.~(2.13) are used, even though neither the 
local terms coming from eliminating the auxiliary fields, nor the nonlocal 
terms arising from graviton and gravitino exchange, is separately 
supersymmetry invariant.  Equation (5.1) gives the leading 
order effects of supergravity coupling on the dynamics of the supersymmetric 
matter fields which give rise to the multiplet of currents, and can 
be used as a starting point for studying whether the  
supergravitational coupling 
can lead to dynamical symmetry breaking in the matter field dynamics.  
This topic will be the subject of a future investigation.  Explicit 
expressions for the components of the current and anomaly multiplets, in 
the notation used here, are given in Appendices B and C for the Wess-Zumino 
and supersymmetric Yang-Mills models, respectively.  

Throughout this paper, we have used the notational conventions for the 
current multiplet employed in the text of West [1].   For completeness, 
we remark that in the notation 
employed in the text of Weinberg [1], Eq.~(5.1) reads 
$$\eqalign{
\kappa^{-2}S_{\rm eff}=&\int d^4x \left[-{3\over 16} 
{\cal R}_{\mu} {\cal R}^{\mu}
+{3\over 4} ({\cal M}^2+{\cal N}^2)\right]\cr
+&\int d^4x d^4y \left[ {1\over 4}T^{\nu\tau}(x) 
(\eta_{\nu\alpha}\eta_{\tau\beta}+ \eta_{\nu\beta}\eta_{\tau\alpha}
-\eta_{\nu\tau}\eta_{\alpha\beta})\Delta_F(x-y) 
T^{\alpha\beta}(y)\right.\cr
-&\left.{1\over 8}  \overline{S}_{\tau}(x) 
\left(\eta^{\tau\nu} \gamma \cdot \partial_{x}+{1\over 2} \gamma^{\tau} 
\gamma \cdot \partial_{x} \gamma^{\nu}\right) \Delta_F(x-y) S_{\nu}(y)
\right]~~~.\cr
}\eqno(5.2)$$
\bigskip
\centerline{\bf Acknowledgments}
This work was supported in part by the Department of Energy under
Grant \#DE--FG02--90ER40542, and benefited from the hospitality 
of the Aspen Center for Physics.  The author wishes to acknowledge 
helpful conversations or email correspondence with Jon Bagger, Eric D'Hoker, 
Dan Freedman, Jim Hartle, Roman Jackiw, Savdeep Sethi, and Ed Witten.

\centerline{\bf Appendix A. ~~ Metric and Gamma Matrix Conventions}

We work with the metric convention 
$$\eta_{\mu\nu}={\rm diag}(-1,1,1,1)~~~,\eqno(A.1)$$
and we use the usual summation convention that repeated Greek indices are 
summed from 0 to 3, together with the abbreviations for four vector inner 
products 
$$\eqalign{
(a_{\mu})^2=&a_{\mu}a^{\mu}~~~,\cr
a \cdot b =& a_{\mu}b^{\mu}~~~,\cr
\square=&\partial_{\mu}\partial^{\mu}~~~.\cr
}\eqno(A.2)$$
Our convention for the four index antisymmetric tensor is 
$$\epsilon^{0123}=-\epsilon_{0123}=1~~~.\eqno(A.3)$$
Our Dirac gamma matrices $\gamma_{\mu}$ obey the anticommutation relations 
$$\{\gamma_{\mu},\gamma_{\nu}\}=2\eta_{\mu\nu}~~~,\eqno(A.4)$$
and we define 
$$\eqalign{
\hat{\gamma}^0=&i\gamma^0 ~~~,\cr
\gamma_5=&i\gamma^1\gamma^2\gamma^3\gamma^0
=i\gamma_0\gamma_1\gamma_2\gamma_3  ~~~,\cr
\sigma_{\mu\nu}=&{1\over 2}[\gamma_{\mu},\gamma_{\nu}]~~~.
}\eqno(A.5)$$
In carrying out the  gamma matrix algebra, we have found it convenient 
to use Majorana representation gamma matrices for which 
$\hat{\gamma}^0 \gamma^{\mu T} \hat{\gamma}^0=-\gamma^{\mu}$, 
with the superscript $T$ denoting 
the transpose.  A convenient explicit representation for these gamma 
matrices in terms of Pauli matrices $\rho_{1,2,3}$ 
and $\tau_{1,2,3}$ is
$$\eqalign{
\gamma^0=&-\gamma_0=-i\rho_2\tau_1~~~,\cr
\hat{\gamma}^0=&\rho_2\tau_1~~~,\cr
\gamma^1=&\gamma_1=\rho_3~~~,\cr
\gamma^2=&\gamma_2=-\rho_2\tau_2~~~,\cr
\gamma^3=&\gamma_3=-\rho_1~~~,\cr
\gamma_5=&-\rho_2\tau_3~~~.\cr
}\eqno(A.6)$$
With Majorana representation gamma matrices the role of the adjoint is 
generally played by the transpose; in particular, we have 
$$\hat{\gamma}^0 (\gamma^{\tau} \gamma \cdot \overleftarrow{\partial} 
\gamma^{\mu})^T \hat{\gamma}^0= -\gamma^{\mu} \gamma \cdot 
\overleftarrow{\partial}  
\gamma^{\tau}~~~,\eqno(A.7)$$
which becomes $\gamma^{\mu} \gamma \cdot \partial \gamma^{\tau}$ after  
integration by parts in the derivation of Eq.~(4.6), 
and the conjugate Rarita-Schwinger spinor $\overline{\psi}^{\mu}$ is 
related to $\psi^{\mu}$ by 
$$\overline{\psi}^{\mu}=\psi^{\mu T} \hat{\gamma}^0~~~. \eqno(A.8)$$

\centerline{\bf Appendix B. ~~ Currents and anomalies 
in the Wess-Zumino model}
We summarize here the formulas for the current and anomaly multiplets in 
the Wess-Zumino model [1].  The action for this model is
$$\eqalign{
S=&\int d^4x {\cal L} \cr
=&\int d^4x\left[-{1\over 2}(\partial_{\mu}A)^2 -{1\over 2}
(\partial_{\mu}B)^2-{1\over 2} \overline{\chi}\gamma \cdot \partial \chi
+{1\over 2} (F^2+G^2) \right.\cr
-&\left.m\left(AF+BG-{1\over 2} \overline{\chi} \chi\right)-\lambda
[(A^2-B^2)F+2GAB-\overline{\chi}(A-i\gamma_5 B)\chi]\right]~~~,\cr
}\eqno(B.1)$$
with $\chi$ a Majorana spinor for which $\overline{\chi}=
\chi^T \hat{\gamma}^0$,  with the superscript $T$ 
again denoting the transpose.  

The Euler-Lagrange equations for this action give the equations of motion
$$\eqalign{
\square A=&mF+\lambda(2AF+2BG-\overline{\chi}\chi)~~~,\cr
\square B=&mG+\lambda(-2BF+2AG+i\overline{\chi}\gamma_5 \chi)~~~,\cr
\gamma \cdot \partial \chi=&[m+2 \lambda(A-i\gamma_5 B)]\chi~~~,\cr
-\overline{\chi}\gamma\cdot\overleftarrow{\partial}=&
\overline{\chi}[m+2\lambda(A-i\gamma_5 B)]~~~,\cr
  F=&mA+\lambda(A^2-B^2)~~~,\cr
G=&mB+2 \lambda A B~~~.\cr
}\eqno(B.2)$$
The action of Eq.~(B.1) is 
invariant under the supersymmetry transformations
$$\eqalign{
\delta A=&\overline{\epsilon}\chi  ~~~,\cr
\delta B=&i \overline{\epsilon} \gamma_5 \chi ~~~,\cr
\delta \chi=&[F + i\gamma_5 G + \gamma \cdot 
\partial(A+i\gamma_5 B)] \epsilon~~~,\cr
\delta \overline{\chi}=&\overline{\epsilon}[F + i\gamma_5 G 
-\gamma \cdot \partial (A-i \gamma_5 B)]~~~,\cr
\delta F=&\overline{\epsilon} \gamma \cdot \partial \chi~~~,\cr
\delta G=&i \overline{\epsilon} \gamma_5 \gamma \cdot \partial \chi~~~.
}\eqno(B.3)$$

The current multiplet in the Wess-Zumino model, which obeys the variations 
of Eq.~(2.7), is given by 
$$\eqalign{
\theta_{\mu\nu}=&-\partial_{\mu}A\partial_{\nu}A
-\partial_{\mu}B\partial_{\nu}B
+{1\over 8}[
\overline{\chi}\gamma_{\mu}\overleftarrow{\partial}_{\nu}\chi
-\overline{\chi}\gamma_{\mu}\partial_{\nu} \chi
+\overline{\chi}\gamma_{\nu}\overleftarrow{\partial}_{\mu}\chi
-\overline{\chi}\gamma_{\nu}\partial_{\mu} \chi]  \cr
-&\eta_{\mu\nu} {\cal L} +{1\over 6}(\partial_{\mu}\partial_{\nu}
-\square \eta_{\mu\nu})(A^2+B^2)~~~,\cr
j_{\mu}=&[-\gamma \cdot \partial(A-i\gamma_5 B) +F+i\gamma_5 G]
\gamma_{\mu}\chi-{2 \over 3}\sigma_{\mu\nu}\partial^{\nu}[(A+i\gamma_5 B)
\chi]~~~,\cr
\overline{j}_{\mu}=&-\overline{\chi} \gamma_{\mu}[\gamma \cdot \partial 
(A + i\gamma_5 B) +F +i\gamma_5 G] +{2 \over 3} \partial^{\nu}
[\overline{\chi}(A+i\gamma_5 B)] \sigma_{\mu\nu}~~~,\cr
j_{\mu}^{(5)}=&-{2 \over 3}\left(B \partial_{\mu}A-A \partial_{\mu}B
+{1\over 4}i \overline{\chi}\gamma_5\gamma_{\mu}\chi\right)~~~,\cr
P=&m(A^2-B^2)~~~,\cr
Q=&2mAB~~~.\cr
}\eqno(B.4)$$
The corresponding anomaly  multiplet, which obeys the variations of 
Eq.~(2.8), is given by
$$\eqalign{
\theta_{\mu}^{\mu}=&-{1\over 2}m\overline{\chi}\chi
+m(AF+BG)~~~,\cr
\gamma \cdot j=&2m(A-i\gamma_5 B) \chi~~~,\cr
\overline{j}\cdot \gamma=&-\overline{\chi}2m(A-i\gamma_5 B)~~~,\cr
\partial \cdot j^{(5)}=&-{2 \over 3}m\left({1\over 2} i \overline{\chi}
\gamma_5 \chi + AG-BF \right)~~~.\cr
}\eqno(B.5)$$

\bigskip
\centerline{\bf Appendix C. ~~ Currents and anomalies in the supersymmetric 
Yang-Mills model}
We summarize here the formulas for the current and anomaly multiplets
in the supersymmetric Yang-Mills model [1].  The action for this model is
$$\eqalign{
S=&\int d^4x {\cal L} \cr
=&{\rm Tr}\left[{1 \over 4 g^2} F_{\mu\nu}F^{\mu\nu}-{
1\over 2} \overline{\chi} \gamma^{\mu}{\cal D}_{\mu} \chi +
{1\over 2} D^2\right]~~~,\cr
}\eqno(C.1)$$
with the field strength $F_{\mu\nu}$ and the covariant derivative 
${\cal D}_{\mu}$ defined by 
$$\eqalign{F_{\mu\nu}=&\partial_{\mu}A_{\nu}-\partial_{\nu}A_{\mu}
+[A_{\mu},A_{\nu}]~~~,\cr
{\cal D}_{\mu}{\cal O}=&\partial_{\mu}{\cal O}+[A_{\mu},{\cal O}]~~~,\cr
}\eqno(C.2)$$
and with $\chi$ again a Majorana spinor.  
Note that we have defined both the gauge potential $A_{\mu}$ and the 
field strength $F_{\mu\nu}$ to be anti-self-adjoint, which is why the 
kinetic term for the gauge field in Eq.~(C.1) has the opposite sign to 
that of Eq.~(3.1), where we took the QED gauge potential to be self-adjoint. 
All fields ${\cal O}=A_{\mu},F_{\mu\nu},\chi,D$ appearing in Eq.~(C.1) 
transform according to the adjoint representation of 
a compact Lie group, with the generator expansion 
$${\cal O}=\sum_a {1\over 2} \lambda_a {\cal O}_a~~~.\eqno(C.3)$$ 
The trace with an upper case T is defined by 
$${\rm Tr}=2{\rm tr}~~~,\eqno(C.4)$$ 
with tr the usual trace for which 
$${\rm tr} \lambda_a\lambda_b =2 \delta_{ab}~~~,\eqno(C.5)$$
so that  
$${\rm Tr} {1\over 2}\lambda_a {1\over 2} \lambda_b = \delta_{ab}~~~.
\eqno(C.6)$$

The Euler-Lagrange equations of Eq.~(C.1) imply the equations 
of motion 
$$\eqalign{
\gamma^{\mu}{\cal D}_{\mu}\chi=&0~~~,\cr
\overline{{\cal D}_{\mu}\chi}\gamma^{\mu}=&0~~~,\cr
{\cal D}_{\mu}F^{\mu\nu}=&g^2\overline{\chi} \gamma^{\nu}\chi~~~.\cr
}\eqno(C.7)$$
The action of Eq.~(C.1) is invariant under the supersymmetry
transformations 
$$\eqalign{
\delta A_{\mu}=&ig\overline{\epsilon}\gamma_{\mu}\chi~~~,\cr
\delta \chi=&\left({i\over 2g}\sigma_{\mu\nu}F^{\mu\nu}+i\gamma_5 D\right)
\epsilon~~~,\cr
\delta \overline{\chi}=&\overline{\epsilon}\left(-{i\over 2g} 
\sigma_{\mu\nu}  F^{\mu\nu}+i\gamma_5 D\right)~~~,\cr
\delta D=&i\overline{\epsilon}\gamma_5 \gamma^{\mu}{\cal D}_{\mu} \chi~~~.\cr
}\eqno(C.8)$$

The current multiplet in the supersymmetric Yang-Mills model, which 
obeys the variations of Eq.~(2.7), is given by 
$$\eqalign{
\theta_{\mu\nu}=&-\eta_{\mu\nu} {\cal L} +...\cr
=&{\rm Tr}\left(-\eta_{\mu\nu} {1\over 4 g^2} F_{\alpha\beta}F^{\alpha\beta}
+{1\over g^2} F_{\nu}^{~~\alpha}F_{\mu\alpha}  \right.\cr
+&\left.{1\over 8}[
\overline{{\cal D}_{\mu}\chi}\gamma_{\nu}\chi
-\overline{\chi}\gamma_{\nu}{\cal D}_{\mu} \chi
+\overline{{\cal D}_{\nu}\chi}\gamma_{\mu}\chi
-\overline{\chi}\gamma_{\mu}{\cal D}_{\nu} \chi ] \right)~~~,\cr
j_{\mu}=&{\rm Tr}\left(-{i \over 2g}F_{\nu\sigma} \sigma^{\nu\sigma} 
\gamma_{\mu} \chi\right)~~~,\cr
\overline{j}_{\mu}=&{\rm Tr}\left(-{i \over 2g} \overline{\chi} 
\gamma_{\mu} \sigma^{\nu\sigma} F_{\nu\sigma}\right)~~~,\cr
j_{\mu}^{(5)}=&-i {1\over 2} {\rm Tr}( \overline \chi \gamma_{\mu} \gamma_5 
\chi)~~~,\cr
P=&0~,~~~Q=0~~~.\cr
}\eqno(C.9)$$
Since the supersymmetric Yang-Mills model is classically conformally 
invariant, the tree level anomalies are zero.  The anomaly multiplet 
which obeys Eq.~(2.8) arises as a one loop radiative correction [7], and 
is given by 
$$\eqalign{
\theta_{\mu}^{\mu}=&-2f{\rm Tr}\left( {1 \over 4g^2} F_{\mu\nu}F^{\mu\nu}
-{1\over 2} \overline{\chi} \gamma^{\mu}{\cal D}_{\mu} \chi+{1\over 2} D^2
\right)~~~,\cr
\gamma \cdot j=&2f {\rm Tr}\left(-{i\over 2g} \sigma_{\mu\nu}F^{\mu\nu} 
+i\gamma_5 D\right) \chi~~~,\cr
\overline{j}\cdot \gamma=&2f{\rm Tr}\overline{\chi}\left(-{i\over 2g} 
\sigma_{\mu\nu}F^{\mu\nu}-i\gamma_5 D\right)~~~,\cr
\partial \cdot j^{(5)}=&-{2\over 3} f {\rm Tr}\left({1\over 4g^2} 
\epsilon_{\mu\nu\rho\sigma}F^{\mu\nu}F^{\rho\sigma}+{i\over 2} 
\partial^{\mu}(\overline{\chi}\gamma_{\mu}\gamma_5 \chi) \right)~~~,\cr
P=&f {\rm Tr} \overline{\chi} \chi~~~,\cr
Q=&if {\rm Tr} \overline{\chi} \gamma_5 \chi~~~,\cr
}\eqno(C.10)$$
with $f$ related to the beta function of the theory by 
$$f={\beta(g) \over g}~~~.\eqno(C.11)$$

\vfill\eject
\centerline{\bf References}
\bigskip
\noindent
[1]  For excellent surveys of standard supersymmetry and supergravity  
topics, see S. Weinberg, {\it The Quantum Theory of Fields, Volume III 
Supersymmetry}, Cambridge University Press, Cambridge, 2000; P. West, 
{\it Introduction to Supersymmetry and Supergravity}, Extended Second 
Edition, World Scientific, Singapore, 1990.\hfill\break 
\bigskip 
\noindent
[2]  K. Stelle and P. West, Phys. Lett. {\bf B74} (1978), 330; 
S. Ferrara and P. van Nieuwenhuizen, Phys. Lett. {\bf B74} (1978), 333.
\hfill\break
\bigskip
\noindent
[3] S. Ferrara and B. Zumino, Nucl. Phys. {\bf B87} (1975), 207.
\hfill\break
\bigskip
\noindent
[4]  P. West, Ref. 1, gives the current multiplet transformation in the form 
used here in his Eq.~(20.43), which differs by some phases from our  
Eq.~(2.7).  \hfill\break
\bigskip
\noindent
[5]  For an exposition, see S. Weinberg, {\it The Quantum Theory of Fields, 
Volume II Modern Applications}, Cambridge University Press, Cambridge, 1996; 
T.-P. Cheng and L.-F. Li, {\it Gauge Theory of Elementary Particle Physics}, 
Oxford University Press, Oxford, 1988, pp. 250-254.\hfill\break
\bigskip
\noindent
[6] Some formulas relevant for this section are 
given in L. Baulieu, A. Georges, 
and S. Ouvry, Nucl. Phys. {\bf B273} (1986), 366; H. A. Weldon, ``Thermal 
Green Functions in Coordinate Space for Massless Particles of Any Spin,''
hep-ph/0007138.\hfill\break
\bigskip
\noindent
[7] See P. West, Ref. 1, Sec. 20.4, for a discussion and further references. 
Note that Eq.~(C.10) differs by phases from West's Eq.~(20.71).  
\bigskip
\noindent
\bigskip
\noindent
\bigskip
\noindent
\bigskip
\noindent
\bigskip
\noindent
\bigskip
\noindent
\bigskip
\noindent
\bigskip
\noindent
\bigskip
\noindent
\bigskip
\noindent
\bigskip
\noindent
\bigskip
\noindent
\bigskip
\noindent
\bigskip
\noindent
\vfill
\eject
\bigskip
\bye